\documentclass[12pt]{article}
\usepackage[english]{babel}

\usepackage{graphicx,amsmath,amssymb}
\begin{document}

\title{On the interband pairing in doped graphane}
\author{Nikolai Kristoffel$^{1}$\footnote{e-mail address: kolja@fi.tartu.ee}\hspace{0.25cm}and K\"ullike R\"ago$^{2}$}
\date{}
\maketitle
\vspace{-1.1cm}
\begin{center}
$^{1}$Institute of Physics, University of Tartu, 142 Riia Street,
51014 Tartu, Estonia \\
$^{2}$Institute of Physics, University of Tartu, 4 T\"ahe Street,
51010 Tartu, Estonia
\end{center}
\begin{abstract}
An estimation shows that the interband pairing channel between the valence band components of doped graphane
can support a superconducting transition temperature (or a contribution into this expected event) of the order of $100$ K at the coupling strength near 1 eV.
\end{abstract}
\begin{flushleft}
\end{flushleft}
The nomenclature of the compounds families exhibiting perspective superconductivity is extended. It enriches perpetually being motivated by fundamental research and proposals for application. Novel systems show often apparent unconventional pairing mechanisms. A complex multiband electron spectrum near the Fermi energy seems to be characteristic for these systems. Accordingly the interband pairing mechanism \cite{MOSKA1991}-\nocite{ISIA1994} \cite{Caivano2009} wins the popularity. The underlying multiband spectrum can appear as a natural property or to be created by special treatments as doping or external pressure. The construction of artifical structures \cite{BIANCONI1994} must also be included detecting high-$T_{c}$ superconductors. Among the apparent multiband-multigap superconductors first the diboride family \cite{Caivano2009}, \cite{Liu2001} -\nocite{Bianconi2004} \cite{Kristoffel2003} and the class of iron based pnictides \cite{Caivano2009}, \cite{Kuroki2008} -\nocite{Izyumov2008, Eschrig2009} \cite{Ishida2009} must be mentioned. Recent approaches to cuprate superconductivity are also of multiband nature \cite{Micnas2003} -\nocite{Belyavsky2007, Alexandrov2005, Kristoffel2004a} \cite{Kristoffel2008}.

The diverse family of carbon compounds includes superconductors with variable structural cages and disperse transition temperatures
\cite{CastroNeto2009} -\nocite{Charlier2007, Kremer2007} \cite{Gunnarsson1997}. Besides the one-band (conventional) models also multiband schemes have been used as e.g. in the case of fullerites ($A_{x}C_{60}$) \cite{KRISTOFFEL1994} -\nocite{CHOI1994} \cite{Suzuki2003}. Superconductivity can be induced in doped diamond \cite{Baskaran2008} -\nocite{Boeri2006, Fukuyama2008} \cite{Wu2006}. The intercalated atoms (e.g. B) create here an impurity band near the top of the valence band and the system acquires the multiband nature. This effect can be compared \cite{Wu2006} with the cuprates where the doped holes create a polaronic defect band above the valence (oxygen)band or under the bottom of the Hubbard band in the case of electron doping \cite{Kristoffel2004a, Kristoffel2008}. Interband pairing working between itinerant and defect bands has been treated in \cite{Kristoffel2004a, Kristoffel2008} as the basic superconductivity mechanism in doped cuprates. The graphite set includes at present the extreme single layer graphene \cite{CastroNeto2009}. It possesses a Dirac fermion type spectrum on which the Fermi level can be shifted by doping. Graphene superconductivity has been treated using multiband mechanisms \cite{Uchoa2007} -\nocite{Kopnin2008} \cite{Lozovik2009}.

A novel result consists in the discovery and investigation of the hydrogenated graphene \cite{Sofo2007} -\nocite{Elias2009} \cite{Wang2009} named graphane. This 2D insulating compound can be doped by holes from intercalant boron atoms. The chemical potential ($\mu$) falls then into two overlapping p-type valence bands which stem from the $sp^{3}$ type bonding like in diamond. These bands have nearly quadratic dispersion with the common origin at the $\Gamma$ point of the Brillouin zone \cite{Elias2009}. This sceleton looks as a classical model for the proof of the pair-transfer superconductivity mechanism \cite{ISIA1994}.

The enhanced density of states near the Fermi energy, as compared with the diamond, has recently stimulated the authors of \cite{Savini2010} to analyze the possible superconductivity of doped graphane. Basing on the calculation of the phonon spectrum of this compound and using the Eliashberg approach the conclusion has been made that the doped graphane can be expected to be a prototype of  a conventional superconductor with extremely high $T_{c}$.

In the present communication we apply a minimal mean-field model to estimate the interband pair-transfer contribution into the expected doped grap\-hane superconductivity. This coupling characterized by a constant W can posses Coulomb and electron-phonon (interband!) contributions. Intraband processes considered in \cite{Savini2010} are here not included. An analogous simplified investigation for $(\mathrm{AlMg})\mathrm{B}_{2}$ has been made in \cite{Rodriguez-Nunez2008}. We have found, that the interband pairing mechanism alone can serve for doped graphane transition temperatures which reach $\sim 100$ K at the coupling strength near 1 eV.

 The pair-transfer interaction between the valence partners causing the superconductivity is taken in the mean
 field approximation as
\begin{equation}\label{eq1}
H_{i}=\Delta_{a}\sum_{\mathbf{k}}
[a_{\mathbf{k}\uparrow}a_{-\mathbf{k}\downarrow}+a^{+}_{-\mathbf{k}\downarrow}a^{+}_{\mathbf{k}\uparrow}]-
\Delta_{b}\sum_{\mathbf{k}}
[b_{\mathbf{k}\uparrow}b_{-\mathbf{k}\downarrow}+b^{+}_{-\mathbf{k}\downarrow}b^{+}_{\mathbf{k}\uparrow}]\,.
\end{equation}
Here the superconducting gap functions are defined by
\begin{eqnarray}\label{eq2}
  \Delta_{a}=2W\sum_{\mathbf{k}}<b_{\mathbf{k}\uparrow}b_{-\mathbf{k}\downarrow}>\,\nonumber\\
   \Delta_{b}=2W\sum_{\mathbf{k}}<a_{-\mathbf{k}\downarrow}a_{\mathbf{k}\uparrow}>\,.
\end{eqnarray}
These expressions contain anomalous operator averages corresponding to Cooper pairs formed from the particles of the same band.
The index \textit{a} is attributed to the more dispersive partner with smaller density of states ($\rho_{a}$). The joint density of states of the doped graphane is nearly constant in the actual energy region around $\mu<0$ \cite{Savini2010}. Accordingly the 2D densities of states $\rho_{a,b}$ are taken to be constant and a nondispersive interaction constant W will be used.

The full Hamiltonian with the band energies $\varepsilon_{a,b}$ counted form $\mu$ the interaction term (\ref{eq1}) can be simply diagonalized. The resulting system of equations for superconducting gaps reads ($\theta=k_{B}T$)
\begin{eqnarray}\label{eq3}
\Delta_{a}=W\Delta_{b}\sum_{\mathbf{k}}\frac{1}{E_{b}(\mathbf{k})}\tanh\frac{E_{b}(\mathbf{k})}{2\theta}\,\nonumber\\
\Delta_{b}=W\Delta_{a}\sum_{\mathbf{k}}\frac{1}{E_{a}(\mathbf{k})}\tanh\frac{E_{a}(\mathbf{k})}{2\theta}\,.
\end{eqnarray}
Here $E_{a,b}(\mathbf{k})=\pm\sqrt{\varepsilon^{2}_{a,b}(\mathbf{k})+\Delta^{2}_{a,b}}$ are the Bogolyubov quasiparticle energies. The equation for the transition temperature, where both gaps $\Delta_{a,b}$ vanish simultaneously is the following
\begin{eqnarray}\label{eq4}
W^{\,2}\sum_{\mathbf{k}}\frac{1}{\varepsilon_{a}(\mathbf{k})}\tanh\frac{\varepsilon_{a}(\mathbf{k})}{2\theta_{c}}
\sum_{\mathbf{k}}\frac{1}{\varepsilon_{b}(\mathbf{k})}\tanh\frac{\varepsilon_{b}(\mathbf{k})}{2\theta_{c}}=1\,.
\end{eqnarray}
We choose the actual interaction region to spread from zero until the cut-off energy $(-E_{c})$. According to the Fig.1 in \cite{Savini2010} we choose $E_{c}=1.5\,$eV. The com\-mon density of states $\rho=0.25\,$(eV)$^{-1}$ has been divided between
the partners

\newpage
by scanning the spectrum in \cite{Savini2010} with the results $\rho_{a}=0.079\,$(eV)$^{-1}$ and $\rho_{b}=0.171\,$(eV)$^{-1}$.

Performing the integrations in (\ref{eq4}) one finds ($\gamma=1.78$)
\begin{eqnarray}\label{eq5}
W^{\,2}\rho_{a}\rho_{b}\left[\ln\frac{4\gamma^{2}|\mu||E_{c}-\mu|}{(\pi\theta_{c})^{2}}\right]^{2}=1\,
\end{eqnarray}
for the chemical potential not too close to the band edges. The transition temperature can be calculated as
\begin{eqnarray}\label{eq6}
\theta_{c}=\frac{2\gamma}{\pi}\sqrt{|\mu||E_{c}-\mu|}\;\mathrm{exp}\left[-\frac{1}{2|W|\sqrt{\rho_{a}\rho_{b}}} \right]
\end{eqnarray}
independently of the sign of $W$. Note the presence of the electron scale energies in front of the exponent in (\ref{eq6}).

Illustrative calculations show the well-known behavior of superconducting characteristics in a two-band model \cite{ISIA1994}. The dependence of $\theta_{c}$ on the interband pairing strength is shown in Fig.\ref{fig1}. Our main result consists in the expectation that the interband interaction alone can induce in the doped graphane superconducting transition temperatures which reach $\sim 100\,$K at the coupling strength of the order of 1 eV. The expected value of W is unknown, however the estimated order seems to be plausible. In the paper \cite{Rodriguez-Nunez2008} for a MgB$_{2}$  family system W$\sim 2.3\,$eV has been found. For A$_{x}$C$_{60}$ a value about 0.4 eV for W has been obtained \cite{Suzuki2003}, cf. also \cite{Uchoa2007} for graphene. The effectiveness of the pairing depends on the position of the chemical potential being optimal for $\mu=-0.75\,$eV. It is known that at 12.5\% hole doping
$\mu=-0.96\,$eV \cite{Sofo2007}. Our result based on the sole interband pairing contribution supports the "intraband" expectation of \cite{Savini2010} about possible high superconducting transition temperatures in doped graphane.

The doping dependences of $\theta_{c}$ on $\mu$ for various coupling strengths are illustrated in Fig.\ref{fig2}. This typical dome form reflects the change of the interband resonance conditions. According to \cite{Kristoffel2004a, Kristoffel2008} in the case of cuprates this effect determines the essential behavior of T$_{c}$ on the whole phase diagram. An uphill of T$_{c}$ with doping leaving the insulating state appears. On contrary, in \cite{Savini2010} a nearly constant T$_{c}$ up to 10\% doping has been found. The behavior for small dopings (p\,$<$\,1\%) remains open in \cite{Savini2010}. Seemingly a remarkable contribution of the interband channel must affect this invariability.

The zero-temperature superconducting gaps can be calculated from the system (cf.\cite{ISIA1994})
\begin{eqnarray}
\Delta_{a}(0)=2\sqrt{|\mu||E_{c}-\mu|}\;\mathrm{exp}\left[-\frac{\Delta_{b}(0)}{2|W|\rho_{a}\Delta_{a}(0)}\right]\,\nonumber\\
\Delta_{b}(0)=2\sqrt{|\mu||E_{c}-\mu|}\;\mathrm{exp}\left[-\frac{\Delta_{a}(0)}{2|W|\rho_{b}\Delta_{b}(0)}\right]\,
\end{eqnarray}
and show a dependence on $\mu$ like in Fig.\ref{fig2}. At this for the characteristic ratios $2\Delta_{a}(0)/\theta_{c}=6.86$
and $2\Delta_{b}(0)/\theta_{c}=4.43$ brake the BCS universality; $\Delta_{a}(0)/\Delta_{b}(0)=1.55$ (independently of $\mu$; $W=1\,$eV).

The discovery of the superconductivity in doped graphane can be expected with a contribution from the interband pairing.
\vspace{1cm}

This work was supported by Estonian Science Foundation grant No. 7296.\\
We are thankful to P. Rubin for discussion.

\bibliography{artikliref2}

Figure 1: Caption: The transition temperature vs the interband pairing strength.\\
Figure 2: Caption: The transition temperature dome on the chemical potential (doping) scale.
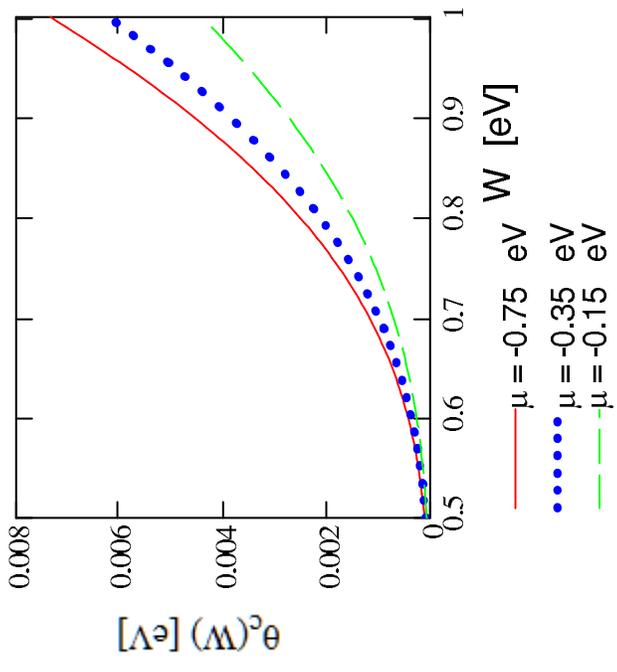
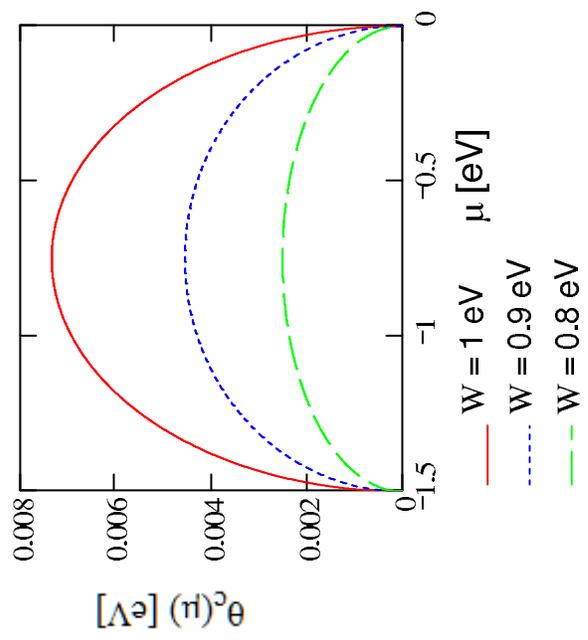
\end{document}